\documentclass[letterpaper,twocolumn,10pt]{article}
\usepackage{usenix,epsfig,endnotes}
\pdfoutput=1
\usepackage{amsmath,amsfonts,amssymb}
\usepackage{mathtools}
\usepackage{bm}
\usepackage{xspace}
\usepackage{booktabs}
\usepackage[sort&compress,square,numbers]{natbib}
\usepackage{multirow}
\usepackage{tabularx}
\usepackage[table]{xcolor}
\usepackage[font={footnotesize}]{caption}

\usepackage{amsthm}
\usepackage{mdframed}
\usepackage{microtype}
\usepackage{array}
\usepackage{adjustbox}
\usepackage{graphicx}
\usepackage{amssymb}
\usepackage{wasysym}
\usepackage{floatrow}
\newcolumntype{Y}{>{\raggedleft\arraybackslash}X}

\newcolumntype{R}[2]{%
  >{\adjustbox{angle=#1,lap=\width-(#2)}\bgroup}%
  l%
  <{\egroup}%
}
\newcommand{\head}[1]{\textnormal{\textbf{#1}}}
\newcommand{\normal}[1]{\multicolumn{1}{l}{#1}}

\usepackage{cite}
\usepackage{url}

\newcommand{\x}{\ensuremath{x}\xspace}
\newcommand{\wm}{\ensuremath{\tilde{x}}\xspace}
\newcommand{\y}{\ensuremath{y}\xspace}
\newcommand{\w}{\ensuremath{w}\xspace}
\newcommand{\transp}[1]{\ensuremath{#1^\intercal}}
\newcommand{\z}{\ensuremath{\tilde{x}}\xspace}

\begin{document}
\date{}

\title{\Large \bf Fraternal Twins: Unifying Attacks on\\
Machine Learning
	and Digital Watermarking}

\author{
	Erwin Quiring, Daniel Arp and Konrad Rieck\\[3mm]
	\begin{minipage}{7cm}
		\centering \it
		Technische Universit\"at Braunschweig\\
		Brunswick, Germany
	\end{minipage}
}

\maketitle

\subsection*{Abstract}
Machine learning is increasingly used in security-critical
applications, such as autonomous driving, face recognition and malware
detection. Most learning methods, however, have not been designed with
security in mind and thus are vulnerable to different types of
attacks. This problem has motivated the research field of
\emph{adversarial machine learning} that is concerned with attacking
and defending learning methods.  Concurrently, a different line of
research has tackled a very similar problem: In \emph{digital
  watermarking} information are embedded in a signal in the presence
of an adversary. As a consequence, this research field has also
extensively studied techniques for attacking and defending
watermarking methods.

The two research communities have worked in parallel so far,
unnoticeably developing similar attack and defense strategies.  This
paper is a first effort to bring these communities together. To this
end, we present a unified notation of black-box attacks against machine
learning
and watermarking that reveals the similarity of both settings. To
demonstrate the efficacy of this unified view, we apply concepts from
watermarking to machine learning and vice versa. We show that
countermeasures from watermarking can mitigate recent model-extraction
attacks and, similarly, that techniques for hardening machine learning
can fend off oracle attacks against watermarks. Our work provides a
conceptual link between two research fields and thereby opens novel
directions for improving the security of both, machine learning and
digital watermarking.

\section{Introduction}\label{sec:introduction}

In the last years, machine learning has become the tool of choice in
many areas of engineering. Learning methods are thus not only applied
in classic settings, such as speech and handwriting recognition, but
increasingly operate at the core of security-critical
applications. For example, self-driving cars make use of deep learning
for recognizing objects and street signs~\citep[e.g.,][]{LevAskBek+11,
  ZhuLiaZha+16}. Similarly, systems for surveillance and access
control often build on machine learning methods for identifying faces
and persons~\citep[e.g.][]{TaiYanRan+14, SchKalPhi15}. Finally,
several detection systems for malicious software integrate learning
methods for analyzing data more effectively~\citep[e.g.,][]{
  LiaYuaWanLi+16, KolLivZorSei12, KapShoCovKruVig13}.

Machine learning, however, has originally not been designed with
security in mind. Many learning methods suffer from vulnerabilities
that enable an adversary to thwart their successful
application---either during the training or prediction phase.  This
problem has motivated the research field of \emph{adversarial machine
  learning} which is concerned with the theory and practice of
learning in an adversarial environment~\citep{HuaJosNelRubTyg11,
  LowMee05b, PapMcSin+16}.  As part of this research, several attacks
and defenses have been proposed, for example, for poisoning support
vector machines \citep{BigNelLas12, BigNelLas11}, crafting adversarial
samples against neural networks~\citep{PapMcDGoo+16,
  PapMcDJhaFreCelSwa16, PapMcDWuJhaSwa16} or stealing models from
decision trees~\citep{TraZhaJuel+16}.

Concurrently to adversarial machine learning, a different line of
research has faced very similar problems: In \emph{digital
  watermarking} information is embedded in a signal, such as an
image, in the presence of an adversary~\citep{CoxMilBlo+02,
  ProHon02}. This adversary seeks to extract or remove the information
from the signal, thereby reversing the watermarking process and
obtaining an unmarked copy of the signal, for example, for illegally
distributing copyrighted content. As a consequence, methods for
digital watermarking naturally operate in an adversarial environment
and several types of attacks and defenses have been proposed for
watermarking methods, such as sensitivity and oracle
attacks~\citep[e.g.,][]{CoxLin97, ComPerPer06, FurBas08,
  BarComPerTon14}.

\begin{figure}[htbp]
\centering
\vspace*{5pt}
\includegraphics[]{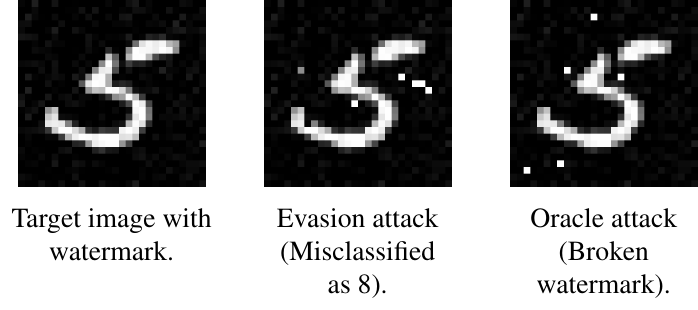}
\caption{Examples of attacks against machine learning
                and digital watermarking. Middle: the target is
                modified, such that it is misclassified as 8. Right:
                the target is modified, such that the watermark is
                destroyed. }
\label{fig:introexample}
\end{figure}

Unfortunately, the two research communities have worked in parallel so
far and unnoticeably developed similar attack and defense strategies.
To illustrate this similarity, let us consider the simplified attacks
shown in Figure~\ref{fig:introexample}: The middle plot corresponds to
an \emph{evasion attack} against a learning method, similar to the
attacks proposed by Papernot et al.~\citep{PapMcDGoo+16,
  PapMcDJhaFreCelSwa16}. A few pixels of the target image have been
carefully manipulated, such that the digit~5 is misclassified as~8. By
contrast, the right plot shows an \emph{oracle attack} against a
watermarking method, similar to the attacks developed by
Westfeld~\citep{Wes08} and Cox \& Linnartz~\citep{CoxLin97}. Again, a
few pixels have been changed; this time however to render the
watermark unreadable in the target image.

\vspace{0.67em}

While both attacks address different goals, the underlying attack
strategy is surprisingly similar. In fact, both attacks aim at
minimally modifying the target, such that a decision boundary is
crossed. In the case of machine learning, this boundary separates
different classes, such as the digits. In the case of digital
watermarking, the boundary discriminates watermarked from unmarked
signals. Although this example illustrates only a single attack type,
it becomes apparent that there is a conceptual similarity between
learning and watermarking attacks.

In this paper, we strive for bringing these two research fields
together and systematically study the similarities of learning and
watermarking methods %
under an adversary's presence with black-box access.
To this end, we
introduce a unified notation for attacks against learning and
watermarking methods, which enables us to reason about their inner
workings and abstract from the concrete attack setting.  This unified
view allows for transferring concepts from machine learning to digital
watermarking and vice versa. As a result, we are able to apply
defenses developed for watermarks to learning methods as well as
transferring machine learning defenses to digital watermarking.

We empirically demonstrate the efficacy of this unified view in two
case studies. First, we show that stateful defenses from digital
watermarking can effectively mitigate model-extraction attacks against
decision trees~\citep{TraZhaJuel+16}. Second, we show that techniques
for hardening machine learning with classifier diversity
\citep{BigCorHeChaGiaYeuRol15} can be successfully applied to block
oracle attacks against watermarks. In addition, we provide further
examples of attacks and defenses, transferable between the research
fields. By doing so, we establish several links between the two
research fields and identify novel directions for improving the
security of both, machine learning and digital watermarking.

In summary, we make the following major contributions in this paper:
\vspace{2pt}
\begin{itemize}
\setlength{\itemsep}{2pt}

\item \emph{Machine learning meets digital watermarking.}  We present
  a novel view on black-box attacks against learning and watermarking 
  methods that exposes previously unknown similarities between both 
  research fields.
\item \emph{Transfer of attacks and defenses.} Our unified view
  enables transferring concepts from machine learning to digital
  watermarking and vice versa, giving rise to novel attacks and
  defenses.
\item \emph{Case studies with two novel defenses.}
  We present and evaluate two novel defenses that are derived from our
  unified view and mitigate model-extraction attacks and oracle
  attacks, respectively.
\end{itemize}
\vspace{2pt}

The rest of this paper is organized as follows: In
Section~\ref{sec:background} we review the background of adversarial
machine learning and digital watermarking. We introduce our unified
view on both research fields in Section \ref{sec:uni} and present case
studies with defenses in Section~\ref{sec:demo}. We discuss the
implications of our work in Section~\ref{sec:discussion} and conclude
in Section~\ref{sec:conclusion}.

\section{Background}\label{sec:background}

Whenever machine learning or digital watermarking are applied in
security-critical applications, one needs to account for the presence
of an attacker. This adversary may try to attack the
learning/watermarking process and thereby impact the confidentiality,
integrity and availability of the application.  This section provides
a basic introduction to the motivation and threat scenarios in
\textit{machine learning} and \textit{digital watermarking}, before
Section~\ref{sec:uni} systematizes them under a common notation.  A
reader familiar with one of the two fields may directly proceed to
Section~\ref{sec:uni}.

\subsection{Adversarial Machine Learning}
Machine learning has become an integral part of many applications in
computer science and engineering, ranging from handwriting recognition
to autonomous driving. The success of machine learning methods is
rooted in its capability to automatically infer patterns and relations
from large amounts of data \citep[see][]{DudHarSto01, HasTibFri01}.
However, this inference is usually not robust against attacks and thus
may be disrupted or deceived by an adversary.
These attacks can be roughly categorized into three classes:
\emph{poisoning}, \emph{evasion} and \emph{model extraction}
\citep{PapMcSin+16}. The latter two are the focus of our work, as they
have concrete counterparts in the area of digital watermarking.

\paragraph{Evasion attacks.}
In this attack setting, the adversary attempts to thwart the
prediction of a trained classifier and evade a detection. To this end,
the attacker carefully manipulates characteristics of the data
provided to the classifier to change the predicted class. As a result,
the attack impacts the \emph{integrity} of the prediction. For
example, in the case of spam filtering, the adversary may omit words
from spam emails indicative for unsolicited
content~\citep{LowMee05}. A common variant of this attack type are
\emph{mimicry attacks}, in which the adversary mimics characteristics
of a particular class to hinder a correct prediction
\citep{SonLocStaKerSto07, FogLee06b}. Evasion and mimicry attacks have
been successfully applied against different learning-based systems,
for example in network intrusion detection~\citep{SonLocStaKerSto07,
  FogShaPerKolLee06}, malware detection~\citep{SrnLas14, XuQiEva16,
  GroPapManBac+16} and face recognition~\citep{ShaBhaBauRei+16}.

Depending on the adversary's knowledge about the classifier, evasion
attacks can be conducted in a \emph{black-box} or \emph{white-box}
setting. In the black-box setting, no information about the learning
method and its training data are available and the adversary needs to
guide her attack along the predicted classes of the
classifier~\citep{LowMee05b,WanParSto06,PapMcDGoo+16}. With increasing
knowledge of the method and data, the probability of a successful
evasion rises~\citep{BigCorMai+13}. In such a white-box setting, the
adversary may exploit leaked training data to build a surrogate model
and then determine what feature combinations have the most effect on
prediction.

\paragraph{Model extraction.}
In this attack setting, the adversary actively probes a learning
method and analyzes the returned output to reconstruct the underlying
learning model~\citep{LowMee05b}. This attack, denoted as \emph{model
  extraction} or \emph{model stealing}, impacts the
\emph{confidentiality} of the learning model. It may allow the
adversary to gain insights on the training data as well as obtain a
suitable surrogate model for preparing evasion attacks.

Depending on the output, the adversary also operates in either a
\emph{black-box} or \emph{white-box} setting. If only the predicted
classes are observable, extracting the learning model is more
challenging, whereas if function values are returned or learning
parameters are available the adversary can more quickly approximate
the learning model. As an example, the recent attacks proposed by
Tram{\`e}r et al.~\citep{TraZhaJuel+16} enable reconstructing learning
models from different publicly available machine learning services in
black-box as well as white-box settings. Moreover, model extraction
poses a serious risk to the privacy of users, as the attack may enable
to derive private information from the reconstructed
model~\citep{ShoStrShm16}.

\subsection{Digital Watermarking}\label{subsec:backgroundWat}

Digital watermarking allows for verifying the authenticity of digital
media, like images, music or videos. Digital watermarks are frequently
used for copyright protection and identifying illegally distributed
content~\citep{website:BBCWatermark}. Technically, a watermark is
attached to a medium by embedding a pattern into the signal of the
medium, such that the pattern is \emph{imperceptible} and
\emph{inseparable}. A particular challenge for this embedding is the
robustness of the watermark, which should persist under common media
processing, such as compression and denoising. There exist several
approaches for creating robust watermarks and we refer the reader to
the comprehensive overview provided by Cox et
al.~\citep{CoxMilBlo+02}.

\begin{figure}[htp]
\centering
\vspace{4pt}
\includegraphics[]{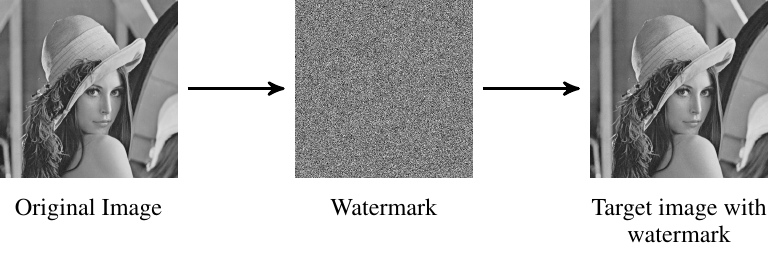}
\vspace{-15pt}
\caption{Example of a digital watermark. A random noise pattern is
  added to the image in the spatial domain. The pattern is not
  observable but detectable.}
\label{fig:realworldexamplewatermarking}
\end{figure}

As an example, Figure~\ref{fig:realworldexamplewatermarking} shows a
simple watermarking scheme where a random pattern is added to the
pixels of an image. The induced changes remain (almost) unnoticeable,
yet the presence of the watermark can be detected by correlating the
watermarked image with the original watermark.
Appendix~\ref{sec:appendixLinearSpecialCase} illustrates this simple
watermarking scheme in more detail.

Similar to machine learning, watermarking methods need to account for
the presence of an adversary and withstand different forms of
attacks~\citep{CoxLin97,FurBas08}. While there exist several attacks
based on information leaks and embedding artifacts that are unique to
digital watermarking \citep[e.g.,][]{CoxMilBlo+02, BasWes09}, we
identify two attack classes that correspond to black-box evasion and
model-extraction attacks.

\paragraph{Oracle attacks.}
In this attack scenario, the adversary has access to a watermark
detector that can be used to check whether a given media sample
contains a watermark or not~\citep{CoxLin97}. Such a detector can be
an online platform verifying the authenticity of images as well as a
media player that implements digital rights management.
Given this detector, the attacker can launch an \emph{oracle attack}
in which she iteratively modifies a watermarked medium until the
watermark is undetectable. The attack thus impacts the \emph{integrity}
of the pattern embedded in the signal.

\begin{figure*}[!t]
\centering
\includegraphics[]{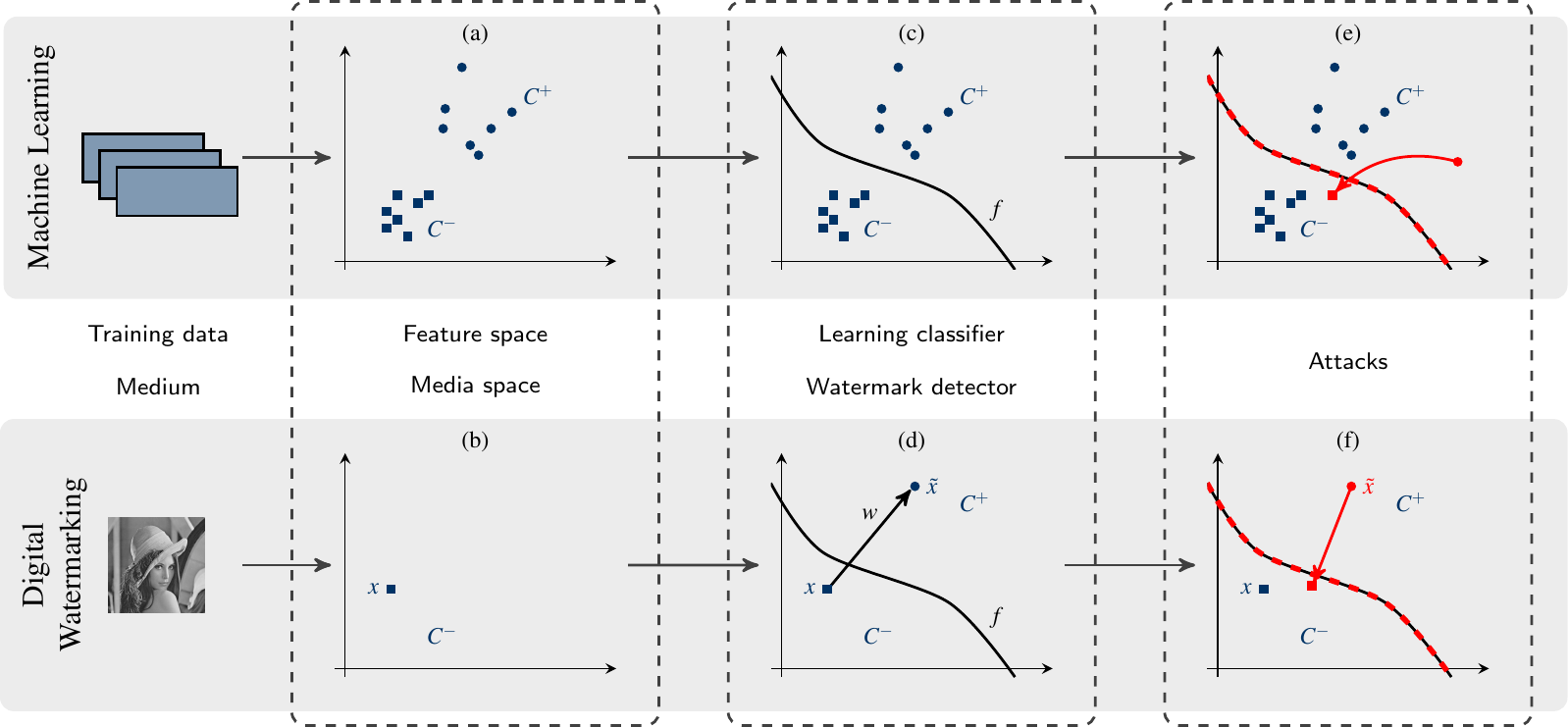}
\vspace{1em}
\caption{A unified view on machine learning and digital
  watermarking. Top: A machine learning setup including a feature
  space, a learning classifier and corresponding attacks. Bottom: A
  watermarking setup including the media space, the watermark detector
  and corresponding attacks.  The red dashed line illustrates model
  extraction/watermark estimation, while the red arrow shows an
  evasion attack/oracle attack.}
\label{fig:unifiedViewFig}
\end{figure*}

While it is trivial to destroy the pattern and the coupled signal, for
example using massive changes to the medium, carefully removing the
watermark while preserving the original signal is a notable challenge.
As a consequence, a large variety of different attack strategies has
been proposed~\citep[e.g.,][]{CoxLin97,Kal98b,ComPerPer06,CraJun07}.
A prominent example is the \emph{Blind Newton Sensitivity Attack},
where no prior knowledge about the detector's decision function is
required and which has been successfully applied against several
watermarking schemes (see Appendix~\ref{sec:bnsa_attack}).

\paragraph{Watermark estimation.} In the second setting, the adversary
also has access to a watermark detector, yet her goal is not only to
remove the watermark from a target medium but to estimate its pattern
\citep{ManTew02a,ChoMou07}. The attack thus impacts the
\emph{confidentiality} of the watermark and not only allows to
perfectly remove the pattern from the signal but also enables forging
the watermark onto arbitrary other data. This \emph{watermark
  estimation} poses a severe threat to watermarking methods, as it may
completely undermine security mechanisms for copyright protection and
access control. However, estimating the pattern embedded in a medium
is difficult and requires a considerable number of queries to the
watermark detector to identify discriminative features in the signal.

\section{Unifying Adversarial Learning \\ and  Digital Watermarking}
\label{sec:uni}
It is evident from the previous section that attacks against learning
and watermarking methods share similarities---an observation that has
surprisingly been overlooked by the two research 
communities~\citep{Barni:2013aa}.
Throughout this section, we systematically identify the similarities
and show that it is possible to transfer knowledge about attacks and
defenses from one field to the other. An overview of this
systematization is presented in Figure~\ref{fig:unifiedViewFig}.
We guide our systematization of machine learning and digital
watermarking along the following four concepts:
\begin{enumerate}
\setlength{\itemsep}{4pt}
\item\emph{Data Representation.} Machine learning and watermarking
  make use of similar data representations, which enables putting
  corresponding learning and detection methods into the same context
  (Section~\ref{subsec:space})

\item\emph{Problem setting.} Watermarking can be seen as a special
  case of a binary classification. Consequently, binary classifiers
  and watermarking techniques tackle a similar problem
  (Section~\ref{subsec:model}).

\item\emph{Attacks.} Due to the similar representation and problem
  setting, attacks overlap between both fields, as we discuss for
  evasion attacks (Section~\ref{subsec:evasion}) and model extraction
  (Section~\ref{subsec:ModelExtraction}).

\item\emph{Defenses.} Defenses developed in one research field often
  fit the corresponding attack in the other field and thus can be
  transferred due to the similar data representation and problem
  setting  (Section~\ref{subsec:defenses}).
\end{enumerate}

In the following, we discuss each of these concepts in more detail,
where we first formalize the concept for machine learning and then
proceed to digital watermarking.

\subsection{Feature Space vs. Media Space}
\label{subsec:space}

\paragraph{Machine learning.}
Learning methods typically operate on a so-called \emph{feature space}
that captures the characteristics of the data to be analyzed and
learned. These features correspond to vectors $\x \in \mathbb{R}^N$
and in the case of \emph{classification} are assigned to a class label
$\y$ that needs to be learned and predicted, such as $C^+$ and $C^-$
in Figure~\ref{fig:unifiedViewFig}(a). Note that feature spaces in
machine learning can also be constructed implicitly, for example using
non-linear maps and kernel functions \citep{SchSmo02,DudHarSto01}.

\paragraph{Digital watermarking.}
Similar to machine learning, watermarking methods operate on a signal
available in some underlying media space, such as the pixels of an
image or the audio waves of a recording.  Without loss of generality,
this signal can be described as a vector $\x \in \mathbb{R}^N$ and
thus the media space corresponds to the feature space used in machine
learning. Note that advanced watermarking schemes often map the signal
to other spaces, such as frequency or random subspace
domains~\citep{CoxMilBlo+02,FurBas08}. Still, the mapped signals can
be described as points in a vector space.\\

Consequently, the feature space of a learning method is closely
related to the media space used in digital watermarking.  The relation
remains unchanged even if a feature mapping is performed, as long as
an implicit vector representation exists.

\subsection{Classifier vs. Watermark Detector}
\label{subsec:model}

\paragraph{Machine learning.}

After embedding the training data into a feature space, the actual
learning process is performed using a learning method, such as a
support vector machine or a neural network. In the case of
classification, this learning method tries to infer functional
dependencies from the training data to separate data
points of different classes. These dependencies are described in a
learning model~$\w$ that parameterizes a decision function
$f_{\w}(\x)$.  Given a vector $\x$ the function $f_{\w}(\x)$ predicts
a class label or a corresponding numerical prediction score.

\paragraph{Digital watermarking.}
The media space in watermarking is divided in two separate subspaces
as depicted in Figure~\ref{fig:unifiedViewFig}(d) where the marked and
unmarked versions of the signal represent the two classes.  Note that
a robust watermark should ideally survive image processing steps, such
as compression and denoising. Therefore, the watermark class
implicitly contains variations as well, just as machine learning
captures the variations of samples from a class through its
generalization.

If we denote an unmarked signal as $\x$ and a watermarked signal as
$\wm$, the relation between $\x$ and $\wm$ is given by a parameter
$\w$ that defines the pattern of the watermark.  As a consequence, a
watermark detector also employs a function $f_{\w}(\x)$ to determine
which subspace a signal is in and thus whether it contains the
watermark. Similar to machine learning, the function $f_\w$ may induce
a linear as well as non-linear boundary, such as a
polynomial~\citep{FurMacHur+02} or fractalized
boundary~\citep{ManTew02b}.\\

Altogether, both fields perform a classification and an adversary
faces the same situation: a decision boundary separates two classes
either in feature or media space.  Consequently black-box attacks that
work through input-output observations are quite transferable between
machine learning and digital watermarking. We emphasize that the
boundary does not need to be the same. Our focus lies on the
corresponding attack strategy.  In the following sections, we discuss
this similarity and provide a mapping between machine learning and
watermarking attacks, which lays the ground for transferring defenses
from one field to the other.

\subsection{Evasion Attack vs. Oracle Attack}
\label{subsec:evasion}

As the first attack mapping, we consider the pair of \emph{evasion}
and \emph{oracle} attacks in a black-box setting. In this attack
scenario, an adversary targets the integrity of the classifier's
response by inducing a misclassification from an iteratively collected
set of input-output pairs. This kind of attack has been proposed for
learning-based classifiers as well as watermark detectors.

\paragraph{Machine learning.} In an evasion attack, the adversary
tries to manipulate a sample with minimal changes, such that it is
misclassified by the decision function $f_{\w}$.  Formally, the attack
can thus be described as an optimization problem,
\begin{align}
\text{arg} \underset{t}{\text{ min }} d(t) \text{ s.t. } f_{\w}(\x + t)
= y^* \; , \label{eq:evasion_opt}
\end{align}
where $d(t)$ reflects the necessary changes $t$ on the original sample
$\x$ to achieve the wanted prediction $y^*$.
Depending on the particular output of the learning classifier, the
attacker can run different attack strategies:
\begin{itemize}
\setlength{\itemsep}{4pt}
\item{\emph{Numerical output.}}  In this case, the classifier returns
  a prediction score $f_{\w}(\x)$ and the attacker tries to mislead
  the classifier with as minimal changes as possible. For example, the
  adversary can perform a gradient descent in the direction of the
  decision boundary to determine the features that have the most
  effect on the classification~\citep[][]{BigCorMai+13,
    PapMcDJhaFreCelSwa16}.

\item{\emph{Binary output.}}
  In the second case, the classifier only
  returns the predicted class label. This clearly restricts the
  attacker's capabilities, since not every change of a feature
  influences the classifier's output. Still, an adversary can perform
  a line search through the binary responses to locate the boundary's
  position~\citep{LowMee05b}.  An attacker can also learn a substitute
  model based on a set of queries that approximates the original
  model~\citep{PapMcDGoo+16,PapMcDGoo16b}.  The attack iteratively
  sends new queries in regions where the substitute model is less
  confident. This allows an evasion even with highly non-linear
  models, such as deep neural networks.
\end{itemize}

Depending on the concrete scenario the attacker might also have to
satisfy additional constraints. For instance, it might not be
sufficient to just cross the decision boundary. Instead, the modified
sample also needs to be located inside the distribution of the target
class~\citep{BigCorMai+13}.

\paragraph{Digital watermarking.}
In an oracle attack, an adversary tries to disturb or even remove the
watermark embedded in a medium. The attack setting is closely related
to evasion. Formally, the underlying optimization problem is given by
\begin{align}
  \text{arg} \underset{t}{\text{ min }} d(t) \text{ s.t. } f_{\w}(\wm + t)
  = y^- \; , \label{eq:oracle_opt}
\end{align}
where $d(t)$ reflects the changes $t$ on the watermarked signal $\wm$
and $y^-$ corresponds to no detection. The optimization problem is
identical to the one given in Eq.~\eqref{eq:evasion_opt}, so that the
adversary can apply similar attack strategies.  We can again
categorize these strategies depending on the output returned by the
watermark detector.

\begin{itemize}
\setlength{\itemsep}{4pt}

\item{\emph{Numerical output.}}
  In this case, the watermark detector
  outputs the score $f_{\w}(\wm)$ of the decision function. As for
  evasion, the adversary can perform an attack based on gradient
  descent to remove the watermark~$\wm$ from the image with as little
  changes as possible~\citep{CoxLin97}.

\item{\emph{Binary output.}}
  In this setting, the adversary has only
  access to the binary output of the watermark detector.  As a
  watermark detector usually does not need to return more information
  than necessary, the watermarking literature generally focuses on
  this setting.  Similar to the evasion case, it is possible to
  perform a line search to locate the decision boundary and remove the
  watermark~\citep[e.g.,][]{ComPerPer06}.
  We present a novel defense against this type of attack in
  Section~\ref{sec:demo} which is inspired by concepts from
  adversarial machine learning.
\end{itemize}

Due to the equivalent objectives in Eq.~\eqref{eq:evasion_opt} and
\eqref{eq:oracle_opt}, attack strategies from machine learning are
transferable to watermarks and vice versa. Take, for instance, the
state-of-the-art Blind Newton Sensitivity Attack~\citep{ComPerPer06}
that solves the optimization problem from Eq.~\eqref{eq:oracle_opt} by
performing a gradient descent based on binary outputs.  This makes the
attack also applicable against non-linear learning classifiers.
Appendix~\ref{sec:bnsa_attack} recaps the attack procedure in more
detail. The optimal solution is guaranteed for convex boundaries, but
suitable results are also reported for non-linear watermarking schemes
by following the boundary's envelope~\citep{ComPerPer06,ComPer07}.\\

We conclude that evasion attacks on classifiers and oracle attacks on
watermark detectors share fundamental similarities
in the black-box setting.
The underlying optimization problems are
identical and thus several of the existing attack and defense
strategies can be directly exchanged from one research area to the
other.

\begin{table*}
	\centering
	\footnotesize
	\setlength{\tabcolsep}{14pt}
	\begin{tabular}{lll}
		\toprule[1.5pt]
		\multirow{2}{*}{\head{Defenses}} &
		\multicolumn{2}{c}{\head{Research Field}}
		\\
		& \normal{\head{Adversarial Learning}} &
		\normal{\head{Watermarking}} \\
		\cmidrule(lr){2-3}
		\vspace{-0.7em} & & \\
		\multirow{2}{10em}{Randomization} &
		Random Subspace Method~\citep{BigFumRol08}& Random Subspace
		Method~\citep{VenJak05,ChoMou06}  \\
		& Randomized Ensemble~\citep{KolTeo09,BigFumRol08} & Randomized
		Boundary ~\citep{LinVan98,FurBas08}  \\
		& --- & Union of Watermarks~\citep{FurBas08}  \\
		\vspace{-0.7em} & & \\
		\cmidrule(lr){2-3}
		\vspace{-0.7em} & & \\
		\multirow{3}{*}{Complex Boundary} &
		Non-Linearity~\citep{BigCorHeChaGiaYeuRol15,RusDemBigFumRol16} &
		Non-Linearity~\citep{ManTew02b,FurMacHur+02,FurVenDuh01,FurBas08}
		\\
		& Classifier
		Diversity~\citep{BigCorHeChaGiaYeuRol15} & --- \\
		& --- & Snake Traps~\citep{FurBas08} \\
		\vspace{-0.7em} & & \\
		\cmidrule(lr){2-3}
		\vspace{-0.5em} & & \\
		\multirow{2}{10em}{Stateful Analysis} & --- & Security
		Margin~\citep{BarComPerTon14,TonComPerBar15} \\
		& --- & Line Search
		Detection~\citep{BarComPerTon14} \\
		& --- & Locality-Sensitive Hashing~\citep{Ven05}  \\
		\vspace{-0.75em} & & \\
		\bottomrule[1.5pt]
	\end{tabular}
	\caption{
		Comparison of defense techniques introduced by adversarial
		learning and digital watermarking.}
	\label{table:defenses}
	\vspace{0.0em}
\end{table*}

\subsection{Model Extraction}\label{subsec:ModelExtraction}
As the second attack mapping, we consider the pair of \emph{model
extraction} and \emph{watermark estimation}. In the black-box scenario,
the adversary aims at compromising the confidentiality of a learning
model or digital watermark by sending specifically crafted
objects to a given classifier/detector and observing the respective
binary output.

\paragraph{Machine learning.} Model-extraction attacks center on an
effective strategy for querying a classifier, such that the underlying
model can be reconstructed with few queries.  For instance, Tram\`{e}r
et al.~\citep{TraZhaJuel+16} have recently demonstrated this threat by
stealing models from cloud platforms providing machine learning as a
service. In contrast to evasion, the extraction of the learning model
$\w$ enables the adversary to also reconstruct the decision function
$f_\w$ and to apply it to arbitrary data. We can differentiate two
attack strategies here:
\begin{itemize}
\setlength{\itemsep}{4pt}

\item \emph{Approximation.}  In the first case, an attacker collects a
  number of input-output pairs with queries either scattered over the
  feature space or created
  adaptively~\citep{TraZhaJuel+16,PapMcDGoo+16,PapMcDGoo16b}. These
  observations allow the adversary to learn an own surrogate model.
  While this strategy is easy to implement, it only yields an
  approximation of the original model, which becomes more accurate the
  more observations are conducted.

\item \emph{Reconstruction.}  The localization of points on the
  decision boundary through a line search enables an exact
  reconstruction of the model.  For example, an adversary can
  reconstruct a linear classifier by performing a line search in each
  feature direction from a fixed position~\citep{LowMee05b}. The
  distance from this position to the located boundary leaks the
  respective feature weight in that direction. The extraction against
  non-linear classifiers such as decision trees also exploits
  localized boundary points for
  reconstruction~\citep{TraZhaJuel+16}. We discuss the latter attack
  in more detail in Section~\ref{sec:demo} when presenting a novel
  defense against it, inspired by concepts from digital watermarking.

\end{itemize}

\paragraph{Digital watermarking.}
Watermark estimation represents the counterpart to model
extraction. In this attack scenario, the adversary seeks to
reconstruct the watermark $\w$ from a marked signal $\wm$. If
successful, the adversary is not only capable of perfectly removing
the watermark $\w$ from the signal $\wm$, but also of embedding $\w$
in other signals, thereby effectively creating forgeries.

Similar to the model extraction case, estimation attacks in the
watermarking literature are based on localizing boundary points where
the signal just crosses the detector's decision
boundary~\citep{ManTew02b,ChoMou07}.  A watermark with a linear
boundary, for instance, can be recovered from a linear number of
discovered boundary points. Choubassi and Moulin present a variant of
this estimation attack to find boundary points that reduce the effort
of the subsequent watermark estimation~\citep{ChoMou07}. This approach
already comes very close to the work of Lowd and
Meek~\citep{LowMee05b} from adversarial machine learning.\\

As digital watermarking and machine learning use a decision boundary
to separate inputs, a natural attack strategy consists in localizing
this boundary through queries and then combining the gathered points to
reconstruct the model or watermark.

\subsection{Defenses}
\label{subsec:defenses}

The communities of both research fields have extensively worked on
developing defenses to fend off the attacks presented in the previous
sections. However, it is usually much easier to create an attack that
compromises a security goal, than devising a defense that effectively
stops a class of attacks. As a result, several of the developed
defenses only protect from very specific attacks and it is still an
open question how learning methods and watermark detectors can be
generally protected from the influence of an adversary. In this
section, we provide an overview of current defenses and identify
similarities as well as interesting directions for transferring a
defense strategy from one field to the other (see
Table~\ref{table:defenses}).  We also include defenses from
adversarial learning that were initially presented against informed
attacks, but also work when an adversary acts in a black-box setting.

\paragraph{Randomization.} A simple yet effective strategy to impede
attacks against classifiers and watermark detectors builds on the
introduction of randomness. Several techniques have been proposed in
both fields which add elements of randomization to the learning or
detection process. While these defenses cannot rule out successful
attacks, the induced indeterminism obstructs simple attack strategies
and requires more thorough concepts for evasion or model extraction.

In machine learning, \emph{randomized ensemble learning} has been
proposed for implementing this defense
strategy~\citep{PerGuLee06,BigFumRol08}. Each classifier in an
ensemble is built with a random subset of the training data and the
prediction is retrieved by aggregating the output of all classifiers.
As a consequence, the adversary has to attack different classifiers at
the same time~\citep{BigFumRol08}. Alternatively, the features
selected to train each classifier can be randomized, such that an
adversary cannot be sure whether a specific feature has an influence
on the returned classifier output~\citep{WanParSto06}. This is known
as the \emph{random subspace method} in the machine learning
field. Overall, randomizing the training data and features raises the
bar for all discussed attacks, since the adversary has to spend more
effort into gaining background knowledge on the underlying decision
boundary and model.

Similar techniques have been proposed to defeat attacks on
watermarking detection. In particular, a detector can be hardened by
creating a \emph{randomized region} around the decision boundary where
the detector returns arbitrary outputs~\citep{LinVan98,FurBas08}.
This misleads the inherent line search in attacks that localize the
boundary in this way~\citep{ComPerPer06,ChoMou07}.
Moreover, equivalent to the random subspace method in machine
learning, several works in the field of watermarking propose to
randomly divide the image pixels into subsets and aggregate the
classifier output from each subset~\citep{VenJak05,ChoMou06}.

In addition, the \emph{Broken Arrows} watermarking scheme creates
several watermarks that form a \emph{union of watermarks}.  During
detection, only the watermark with the smallest distance to the
current signal is applied~\citep{FurBas08}. This mitigates the risk
that an adversary could compare multiple images with the same
watermark. This defense has not been applied to learning methods
yet. It would correspond to an ensemble of classifiers where only one
classifier is applied during prediction depending on the input sample.

\begin{figure*}[!t]
\centering
\includegraphics[]{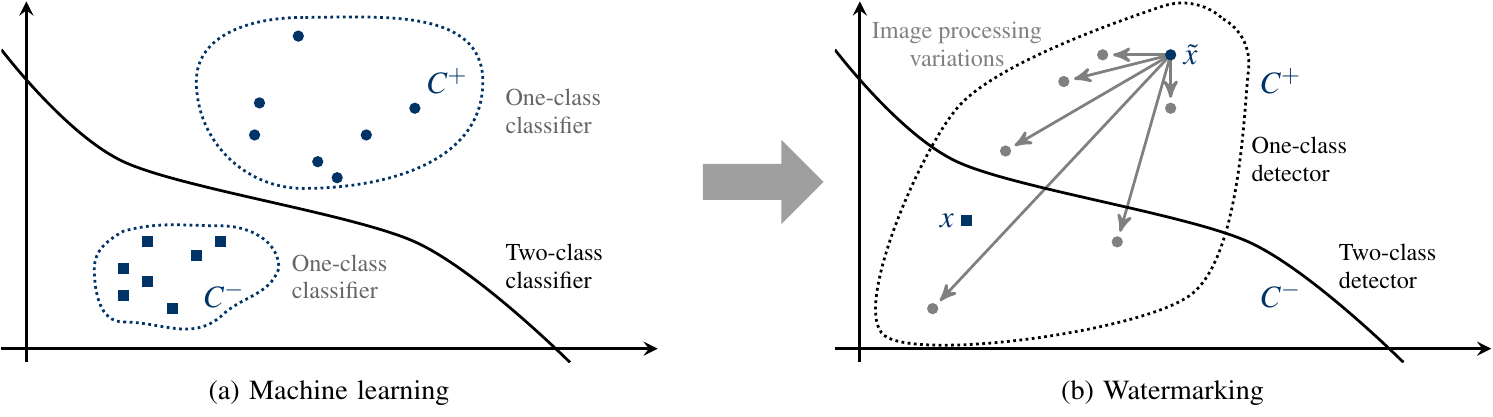}
\caption{Transfer from machine learning to watermarking: The left plot
shows the combination of a two-class and one-class learning model to
build a so-called 1.5-class classifier. The right plot
shows our novel defense for watermarking that also combines a two-class
and one-class detector.}
\label{fig:MLtoWDW}
\end{figure*}
\vspace{1em}

\paragraph{Complex boundary.} Another strategy for obstructing attacks
is the selection of a complex decision boundary. Without sufficient
knowledge on the structure of the boundary, it is difficult to attack
a method and the adversary is required to invest more resources to
circumvent the defense. However, increasing the complexity of a
decision function is not trivial, as a fine-grained boundary can also
lead to overfitting and further possibilities for evasion
\citep[see][]{PapMcDJhaFreCelSwa16}.

Recent work on adversarial machine learning thus proposes to increase
the complexity of the decision boundary but at the same time to
enclose the learned data tightly. In the case of malware detection,
this implies that an evasion attack needs to contain plausible
features of the benign class without losing the malicious
functionality. Russu et al.\ implement this defense strategy using
non-linear kernel functions~\citep{RusDemBigFumRol16}, while Biggio et
al. realize a tighter and more complex boundary through the
combination of two-class and one-class
models~\citep{BigCorHeChaGiaYeuRol15}.  Although invented against
attackers with a surrogate model, these countermeasures also tackle
black-box attacks that need to probe the feature space with queries
outside the training data distribution. We use this strategy in
Section~\ref{subsec:MLtoDW} to address a watermark oracle attack.

Similarly, the watermarking community has examined non-linear
boundaries to defend against oracle and watermark-estimation
attacks~\citep{ManTew02b, FurMacHur+02, FurVenDuh01}. These boundaries
range from polynomial to non-parametric fractals and obstruct e.g.\
attacks that estimate the decision boundary. In addition, Furon and
Bas have introduced small indents called \emph{snake traps} at the
decision boundary in order to stop attacks based on random walks along
the detection region~\citep{FurBas08,CraJun07}.

While all of the defenses listed in Table~\ref{table:defenses} using a
complex boundary render attacks more difficult, it is often not clear
whether they provide protection in the long run. For example,
boundaries based on fractals and snake traps block some of the attacks
presented in Section~\ref{sec:uni}, yet approximations of the decision
boundary are still possible and might be used for successful attacks
\citep{ComPerPer06}.

\paragraph{Stateful analysis.} If the learning method or watermark
detector is outside of the attacker's control, an active defense
strategy becomes possible, in which the defender seeks to identify
sequences of malicious queries. For instance, a cloud service
providing machine learning as a service may monitor incoming queries
for patterns indicative of evasion and model-extraction attacks.

While this concept has not yet been examined in adversarial machine
learning, stateful analysis of queries has been successfully applied
in digital watermarking for detecting oracle and watermark-estimation
attacks~\citep{Ven05,BarComPerTon14,TonComPerBar15}.
These defenses exploit
the fact that an adversary first needs to perform an unusual number of
queries close to the boundary in order to exactly locate its
position. Thus, it is possible to detect attempts to attack the
detector and infer the decision boundary.

Consequently, Table~\ref{table:defenses} shows that stateful defenses
have only been applied to watermarking schemes, providing the
opportunity for constructing novel defenses for learning methods. We
show in a case study in Section~\ref{sec:demo} that model-extraction
attacks can be mitigated with the security margin concept if the
learning system is not under full control of the adversary and it is
possible to monitor incoming queries.

\section{Transfer of Attacks and Defenses}
\label{sec:demo}
We proceed to present two case studies that exemplify how concepts
from one research field can be transferred to the other. As the first
case study, we apply a concept proposed by Biggio et
al.~\citep{BigCorHeChaGiaYeuRol15} for securing machine learning to a
watermark detector. We demonstrate that the resulting detector 
mitigates a state-of-the-art oracle attack. In the second 
case study, we apply the concept of stateful detection to a machine
learning method and show that this combination effectively tackles 
model-extraction attacks against decision trees. While these
case studies focus on two particular defenses, we encourage the
communities to work with each other and therefore summarize further
directions for research in Section~\ref{sec:discussion}.

\subsection{From Machine Learning \\to Watermarking}
\label{subsec:MLtoDW}

In our first case study, we consider a recent defense from machine
learning that increases the complexity of the decision boundary by
combining a two-class and one-class
model~\citep{BigCorHeChaGiaYeuRol15}. Figure~\ref{fig:MLtoWDW}(a)
schematically illustrates the concept of this defense, which
effectively creates a blend between two independent learning methods:

\begin{itemize}
\setlength{\itemsep}{3pt}
\item \emph{Two-class models}. The objective of this learning setting
  is to discriminate objects from two classes. However, unpopulated
  regions, such as the top-left corner in Figure~\ref{fig:MLtoWDW}(a),
  are not excluded from this classification, which leads to a weak
  spot: an attacker can try to evade the classification by creating
  arbitrary samples on the selected subspace of the decision
  function---irrespective of the distribution of the target class.

\item \emph{One-class models.}  In this learning setting, only one
  class is modeled and the decision boundary separates this class from
  the rest of the feature space.  Figure~\ref{fig:MLtoWDW}(a)
  exemplifies the concept by showing the learned boundary around the
  classes. One-class models enable identifying implausible points, as
  they tightly enclose the training data and thereby help to mitigate
  the weak spot of common two-class models.
\end{itemize}

If we combine the decision boundary of both models, we obtain a hybrid
form denoted as \mbox{``1.5-class classifier''}. This classifier
discriminates two classes but also requires these classes to lie
within specific regions of the feature space. As a result, evasion
attacks become more difficult, since an adversary needs to stay within
the one-class regions when moving towards the decision boundary.

\begin{figure}[t]
	\centering
	\includegraphics[]{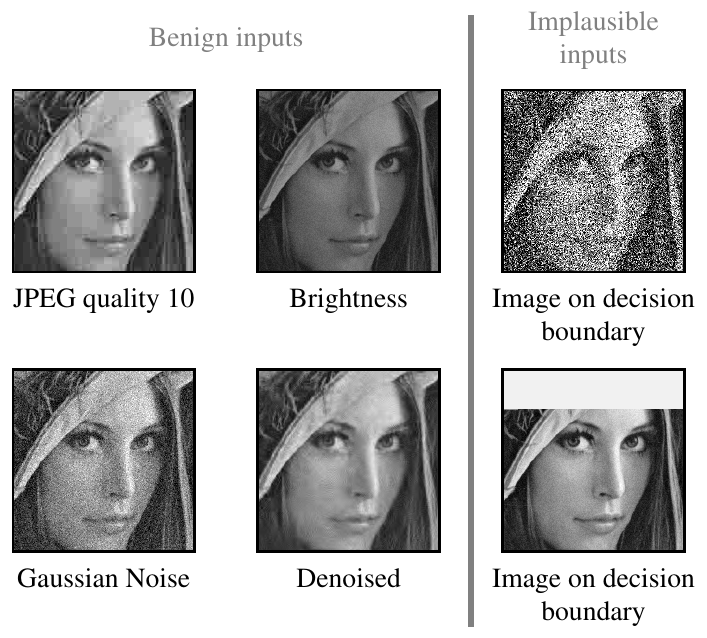}
	\caption{Distortions of the target image. The left four plots
          show plausible image distortions. The right plots depict
          boundary starting positions for an attack where the 
          one-class detector gives an alarm.}
	\label{fig:MltoDW_Examples_Groupplot}
\end{figure}

This simple yet effective idea has been proposed for learning methods
but has not been applied in the context of digital watermarking. In
fact, existing watermarking schemes mainly focus on discriminating
marked from unmarked signals and neglect how these are distributed in
the media space, leading to the same weak spot.
An adversary can therefore exploit the full media space to 
trigger varying reactions to the respective inputs in order to collect 
information about the watermark. Broadly speaking, ''the image does not 
have to look nice`` in an attack~\citep{Wes09}. The so-created points 
do not necessarily resemble a meaningful signal, but the detector still 
accepts these points.
The Blind Newton Sensitivity Attack from Section~\ref{subsec:evasion},
for instance, needs to find a starting position on the decision
boundary. Without further information about the boundary's 
location, an attacker can thus perform a line search in a random 
direction or set pixels to gray iteratively (see
Figure~\ref{fig:MltoDW_Examples_Groupplot} for the resulting images).

Consequently, the defense from the learning community provides us with
a new research direction to tackle oracle attacks in digital
watermarking. Figure~\ref{fig:MLtoWDW}(b) depicts a possible
\textit{1.5-class~watermark detector} that works as follows: The
two-class detector enables us to distinguish unmarked from watermarked
signals, while the one-class detector enables spotting implausible
signals, that is, too far away from reasonable variations.  In
particular, the two-class detector decides on the presence of a
watermark only if the input lies within the one-class region. Outside,
the detector returns random decisions or alternatively blocks access
for subsequent queries from the same source
(see~Section~\ref{subsec:defenses}).

To model plausible signals, the 1.5-class watermark detector is
trained with samples of honest variations of the target image, such as
strong changes of the brightness, compression or denoising.
Figure~\ref{fig:MltoDW_Examples_Groupplot} depicts possible variations
with distortions where the detector should still decide on watermark 
presence.

\paragraph{Experimental setup.}
To demonstrate the practical utility of this novel defense, we conduct
an empirical evaluation with a state-of-the-art oracle attack. Our
dataset for this evaluation consists of images from the publicly
available Dresden Image Database~\citep{Gloe:2010ad}, where 50
uncompressed Adobe Lightroom images from a Nikon~D70 camera are
used. All images are converted to grayscale and cropped to a common
size of $128\times 128$ pixels, that is, $N=16834$ dimensions.

Our experimental procedure is as follows: The watermark embedding and
detection process follow the presented scheme in
Appendix~\ref{sec:appendixLinearSpecialCase} that yields a linear
decision boundary and that represents the two-class detector. To
obtain the respective one-class detector, we create different
variations of the watermarked image $\z$ by applying common image
processing steps, such as noise addition, denoising, JPEG compression
as well as contrast- and brightness variation. We apply
neighborhood-based anomaly detection to define a simple model of
normality.
Given an image $\wm$, this model computes the distance $d$
to the $k$-nearest variation of $\wm$, that is,
\begin{align}
d(\wm) = \frac{1}{k} \sum_{v \in N_{\wm}} \Vert v - \wm \Vert \;
\label{eq:anomaly_detection}
\end{align}
where $N_{\wm}$ are the $k$-nearest neighbors of $\wm$. We mark an
image as implausible if the distance to its $k$-nearest variations
reaches a given threshold $\delta$. For our study, we simply fix $k=3$.
We attack our 1.5-class detector using the well-studied Blind Newton
Sensitivity Attack~\citep{ComPerPer06,ComPer07,BarPerComBar07}
that successfully defeats several existing defenses (see
Appendix~\ref{sec:bnsa_attack}). We perform the attack against each of 
the 50~images and report aggregated results.

\begin{figure}[t]
	\centering
	\includegraphics[]{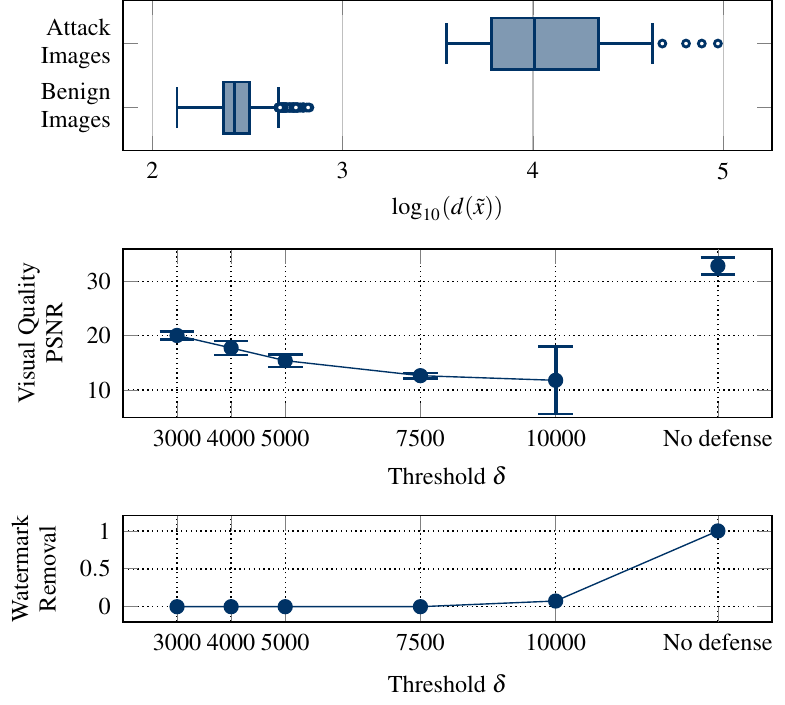}
	\vspace{-1.05em}
	\caption{
	The upper plot compares the average distance from normally 
	distorted and adversely crafted images to a respective one-class 
	model. 
	The middle plot shows the average PSNR of the 50 image outcomes 
	of the attack to the respective original, unwatermarked image as a 
	function of $\delta$. The error bars depict the 
	respective standard deviation. The lower plot depicts the 
	percentage of the attack images where the watermark is not 
	detectable anymore.
	}
	\label{fig:MltoDW_Boxplot}
\end{figure}

\paragraph{Benign vs. attack images.}
We start with a comparison between benign and attack images.  For each
benign image $\z$, we randomly split its set of image variations into
a known partition (75\%) and unknown partition (25\%). We repeat this
procedure 50~times and report the distance of each distorted image
from the unknown partition to the one-class model defined by the known
partition. Figure~\ref{fig:MltoDW_Boxplot}(a) shows the average
distance for each benign image and repetition as well as the attack 
queries per image $\z$.

The distances between benignly distorted and adversely crafted 
images are well separable. Without further information about the 
boundary, an attacker needs to use an arbitrary localized boundary 
position. Thus, the gradient calculation around that position leads to 
a large fraction of queries that do not resemble a meaningful signal 
and exhibit an abnormally high distance. This in turn allows a 
one-class detector to differentiate between benign and attack inputs.

\paragraph{Random Decisions.}

We finally evaluate the impact of random decisions outside the
one-class model. %
Figure~\ref{fig:MltoDW_Boxplot}(b) plots the Peak Signal
To Noise Ratio (PSNR) and Figure~\ref{fig:MltoDW_Boxplot}(c) shows the 
percentage of successfully eliminated watermarks from the
50~final attack outcomes as a function of threshold~$\delta$. 

A threshold between $3000$ and $10000$ successfully 
distorts the attack, such that the watermark is still 
detectable in the final attack outcome and the visual quality decreases 
substantially.
A higher threshold, however, increases the chances that 
the attack queries remain inside the one-class model, so that an 
adversary can render the watermark undetectable with a high visual 
quality again. A smaller threshold increases the risk of false 
positives. Moreover, if the model becomes too tight, the attack 
terminates around the one-class boundary instead. This leads to 
increasing PSNR values and an adversary could start to exploit the 
one-class boundary instead. 
Overall, the results confirm that a suitable threshold strongly 
impacts the attack with regard to visual quality and watermark 
removal, and at the same time reduces the risk of false positives.

We acknowledge that false positives may occur with image distortions 
that are not considered in the one-class model. Yet, the model
represents an additional information source for a watermark detector. 
To the best of 
our knowledge, a similar defense strategy has not been proposed for 
digital watermarking so far and in combination with already existing 
defense mechanisms, we further raise the bar for oracle and 
watermark-estimation attacks.

\begin{figure*}[!t]
	\centering
	\includegraphics[]{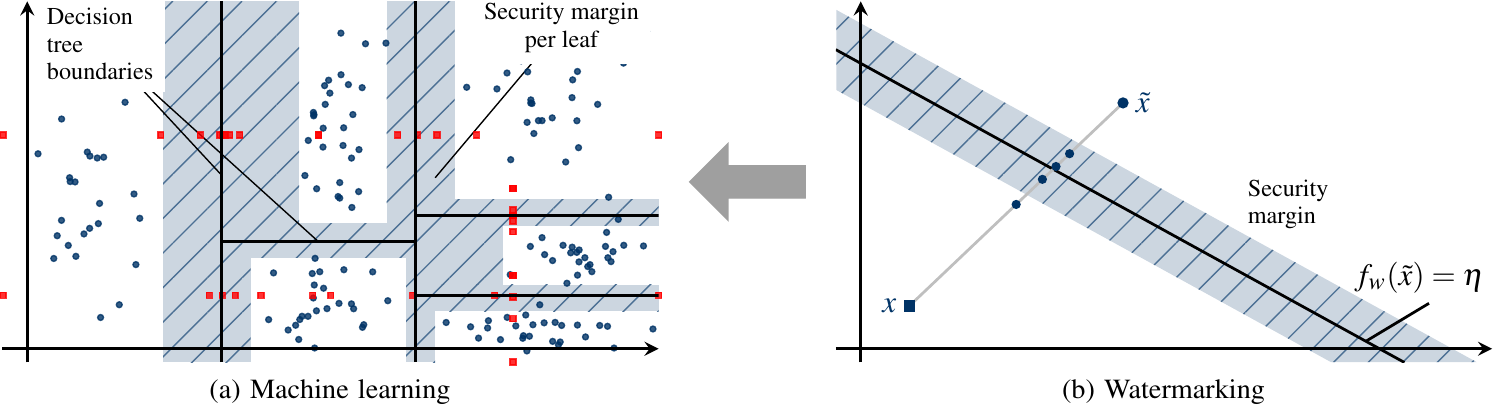}
	\caption{
		Transfer from digital watermarking to machine learning. The
		right plot illustrates the concept of the security margin that
		spots a binary search trying to enclose the boundary. The left
		plot shows its application to a decision tree. The region
		defined by each leaf has a security margin where its width is
		adapted to the training data distribution (circles).
		In contrast, the queries from the tree extraction algorithm
		(squares) underline that the attack needs to operate in the
		security margin to localize the decision boundary.}
	\label{fig:DWtoMLtree}
\end{figure*}

\subsection{From Watermarking\\ to Machine Learning}
\label{subsec:DWtoML}

We proceed with applying concepts from the field of watermarking to
machine learning. In particular, we demonstrate that the concept of a
stateful detector mitigates the risk of model stealing by identifying
sequences of malicious queries. This provides online services offering
machine learning as a service new capabilities for fending off
attacks. To begin with, we shortly summarize the model extraction
proposed by Tram{\`e}r et al.~\citep{TraZhaJuel+16} for a decision
tree and then develop a stateful classifier as an effective
countermeasure. Finally, we present an empirical evaluation to
demonstrate the practical feasibility of this novel defense strategy.

\paragraph{Decision tree extraction.}
Tram{\`e}r et al.~\citep{TraZhaJuel+16} reconstruct decision trees by
performing targeted queries on the APIs provided by the BigML service.
The attack is possible, since the service does not only return the
class label for a submitted query but also a confidence score for a
particular leaf node. This enables an adversary to distinguish between
the leaves.  For each leaf and for each of its features, a recursive
binary search locates the leaf's decision boundary in that
direction. As the binary search covers the whole feature range, other
leaf regions are discovered as well and extracted subsequently. In
this way, an adversary can extract all possible paths of the decision
tree. Note that the binary search needs to fix all features except for
the one of interest, as otherwise the attack may miss a leaf during
the reconstruction.

\begin{table*}
	\centering
	\small
	\begin{tabularx}{0.95\textwidth}{r *{9}{Y}}
		\toprule[1.5pt]
		\multirow{2}{*}{\head{Dataset}} &
		\multicolumn{2}{c}{\head{Original Attack}} &
		\multicolumn{2}{c}{\head{Blocking Defense}} &
		\multicolumn{2}{c}{\head{Random Resp. Defense}} &
		\multicolumn{2}{c}{\head{Adapted Attack}} \\
		& \head{Q}& \head{p}
		& \head{Q}& \head{p} & \head{Q}& \head{p}
		& \head{Q}& \head{p} \\
		\cmidrule(lr){1-1}\cmidrule(lr){2-3}
		\cmidrule(lr){4-5}\cmidrule(lr){6-7}\cmidrule(lr){8-9}
		Iris  & 108 & 1.00 & 38 & 0.09 & * & 0.09 & 4,412 & 1.00
		\\
		Carseats  & 871 & 1.00 & 148 & 0.20 & * & 0.20 & 15,156 &
		0.46 \\
		College  & 2,216 & 1.00 & 244 & 0.10 & * & 0.10 & 8,974 &
		0.08 	\\
		Orange Juice & 4,804 & 1.00 & 846 & 0.20 & * & 0.20 &
		86,354 & 0.48 \\
		Wine Quality & 9,615 & 1.00 & 978 & 0.11 & * & 0.11 &
		37,406 & 0.11 \\
		\bottomrule[1.5pt]
	\end{tabularx}
	\caption{Effectiveness of the Security Margin Defense for different
		attack variations and possible reactions after detecting the
		attack. $Q$ denotes the number of queries, $p$ the percentage
		of
		successfully extracted leaves. Without any defense, the
		original
		extraction algorithm from Tram{\`e}r et al.\ extracts the whole
		tree ($p=1$).
		In contrast, with the security margin, the detector can spot an
		attack
		and block further access before the whole tree is reconstructed
		($p\leqslant0.2$). If the adversary adapts the attack by 
		sending cover
		queries with random values, the attack chances increases, but
		the full
		reconstruction is still not possible for larger datasets.}
	\label{table:evaluation_securitymargin}
\end{table*}

\paragraph{Stateful decision tree.}
As a countermeasure to this attack, we devise a defense that builds on
a successful protection technique from digital
watermarking---a~\emph{stateful detector}~\citep{BarComPerTon14,
  TonComPerBar15}. Figure~\ref{fig:DWtoMLtree}(b) shows the concept as
proposed by Barni et al.~\citep{BarComPerTon14}. A narrow stripe
across the decision boundary determines a \textit{security
  margin}. The detector does not only check for the presence of a
watermark, but simultaneously counts the number of queries falling
inside this margin.  An attacker performing a binary search to enclose
the boundary will necessarily create an unusually large number of
queries in the security margin. The analysis of the input sequences
therefore allows the identification of unusual activity which
mitigates the risk of oracle and watermark-estimation attacks.  The
exact parameters of the security margin are derived through
statistical properties of the decision
function~\citep{BarComPerTon14}.  Although this defense strategy has
been initially designed to protect watermark detectors, we
demonstrate that it can be extended to secure decision trees as well.

Figure~\ref{fig:DWtoMLtree}(a) illustrates the transferred concept where
security margins are added to the boundaries of each tree region.
The width of these margins is determined for each region and feature
dimension separately depending on the statistical distribution of the
data.
Overall, the security margin is defined
alongside the original decision tree and does not require changes to
its implementation. Appendix~\ref{sec:appendixSecurityMargin} provides
more information on the margin's creation process.

When the decision tree returns the predicted class for a query, the
stateful detector checks whether the query falls inside the security
margin. To determine whether the tree is subject to an attack, we
calculate a simple ratio: For each leaf, we count the number of
incoming queries. At the same time, the leaf keeps record of the
queries inside the security margin. We denote by $\varphi$ the ratio
from the security margin queries to the total number of queries,
averaged over all leaves. This ratio is an indicator for the
plausibility of the current input sequence.
Figure~\ref{fig:DWtoMLtree}(a) also shows the typical query sequence
from the tree extraction algorithm (red squared). The adversary has to
work in the margin to localize the decision boundary, in contrast to
the distribution of benign queries.

\paragraph{Experimental setup.}
To evaluate this defense in practice, we use the publicly available
tree-stealing implementation by Tram{\`e}r et al.~\citep{TraZhaJuel+16}.
Table~\ref{tab:dataset} summarizes our used datasets. We divide each
dataset into a training set (50\%) and test set (50\%), where we use
the first for learning a decision tree and calibrating the security
margins. We repeat this process 5~times and present aggregated results
in the following.
The detector assumes an attack if the query ratio $\varphi$ exceeds the
threshold $\tau = 0.3$, estimated by a prior cross-validation. 

\begin{table}[ht]
	\centering
	\footnotesize
	\begin{tabular}{rrrr}
		\toprule[1.5pt]
		\head{Dataset} & \head{Samples} & \head{Features} &
		\head{$\varnothing$ Leaves}
		\\
		\midrule
		Iris & 150 & 4 & 4.6 \\
		Carseats & 400 & 8 & 13.2 \\
		College & 777 & 17 & 18.8 \\
		Orange Juice & 1,070 & 11 & 59.0 \\
		Wine Quality & 1,599 & 11 & 89.4 \\
		\bottomrule[1.5pt]
	\end{tabular}
		\caption{Dataset for evaluation. The number of leaves from the
			learned decision tree are averaged over the repetitions.}
		\label{tab:dataset}
\end{table}
\vspace{-1em}

\paragraph{Defense Evaluation.}
We first examine the security margin under benign and attack queries,
where Table~\ref{table:evaluation_securitymargin} reports the results
for the corresponding experiments. In the first step, we make use of
the test set to simulate the queries of an honest user. In this way,
we can determine the risk of false positives, that is, declaring that
an honest input sequence is malicious.
The final query ratio $\varphi$ after submitting the complete sequence
was not higher than 0.2 in all datasets, so that the stateful detector
does not mark a benign query sequence as attack by mistake.

Next, we run the tree-stealing attack against the learned tree without
and with the security margin defense. In the latter case, we consider
two reactions after that an attack sequence is detected: (a) the
tree blocks further access, (b) the tree returns random
decisions. To determine the knowledge gain by the adversary,
Table~\ref{table:evaluation_securitymargin} reports the percentage of
successfully extracted leaves~$p$. The blocking strategy allows the
tree to block the tree extraction at the very beginning.
With random decisions, the attack's binary search recursively locates
an exponential number of boundaries erroneously. We
stopped the attack after 1~Million queries (marked by *).

\paragraph{Counter-Attack Evaluation.}
As a counter-reaction, an adversary can in turn submit \emph{cover
  queries} outside the security margin so that the query ratio 
  $\varphi$ ideally remains below the threshold. There are, however, 
  two practical
problems. Without knowledge of the training data distribution, the
adversary cannot know where a decision boundary could be located and
thus where the margin could be. Another problem is that the attacker
needs to control the ratio in almost each leaf. It is not sufficient
to send just one fixed well-chosen cover query all the time, since
this query would only affect one leaf.  These two problems complicate
the design of cover queries.

We therefore let the attacker create cover queries by selecting random
values in the range of each feature.
Table~\ref{table:evaluation_securitymargin} shows the performance of 
this adapted attack where an adversary sends 40~cover queries for each 
tree extraction query. Still, the whole tree cannot be extracted. Only 
half of the leaves are extracted before the detector spots
the attack and blocks further access. 

We finally consider a stronger attacker who knows a certain percentage
of the training data.  This is not unrealistic, if an adversary can
make some assumptions about possible training data. The attacker can
make use of the leaked training data as cover
queries. Table~\ref{tab:tabKnowledgeAttacker} summarizes the
percentage of extracted leaves $p$ for varying amounts of known
training data and cover queries. If just 10\% of the data are known,
even 40~cover queries between each attack query do not suffice to
extract the whole tree. However, if the adversary knows more data
points, the cover queries spread over all leaves more equally and the
attack chances start to increase.

Overall, our evaluation demonstrates that the proposed defense can
effectively mitigate the risk of model stealing based on the history
of queries. 
While this defense is only a first step in hindering
model-extraction attacks, we show that the concept of a stateful
analysis brings in a new defense strategy that in combination with
other protections, such as a line search detection as proposed for
watermarking~\citep{BarComPerTon14}, can lower the chances of
reconstructing a model in reasonable time.  As our defense can be
implemented alongside an existing classifier, online services such as
BigML can easily deploy our defense in practice.

\begin{table}[t]
	\scriptsize
	\centering
	\begin{tabular}{rrccccc}
		\toprule[1.5pt]
		\multirow{2}{*}{Dataset}  & Cover &
		\multicolumn{5}{c}{Percentage train. data} \\
		\cmidrule{3-7}
		& Queries & 10 & 20 & 30 & 40 & 50 \\
		\midrule
		\multirow{3}{*}{Iris}
		& 1x & 0.17 & 0.21 & 0.21 & 0.21 & 0.22 \\
		& 5x & 0.64 & 0.85 & 0.89 & 0.92 & 0.94 \\
		& 40x & 0.76 & 0.91 & 0.94 & 0.97 & 1.00 \\
		\midrule
		\multirow{3}{*}{Carseats}
		& 1x & 0.28 & 0.29 & 0.28 & 0.29 & 0.30 \\
		& 5x & 0.39 & 0.60 & 0.69 & 0.82 & 0.89 \\
		& 40x & 0.50 & 0.87 & 0.97 & 1.00 & 1.00 \\
		\midrule
		\multirow{3}{*}{College}
		& 1x & 0.12 & 0.12 & 0.12 & 0.12 & 0.12 \\
		& 5x & 0.17 & 0.26 & 0.28 & 0.29 & 0.32 \\
		& 40x & 0.29 & 0.64 & 0.85 & 0.94 & 1.00 \\
		\midrule
		\multirow{3}{*}{Orange Juice}
		& 1x & 0.28 & 0.29 & 0.29 & 0.29 & 0.29 \\
		& 5x & 0.39 & 0.63 & 0.88 & 0.98 & 0.99 \\
		& 40x & 0.46 & 0.92 & 1.00 & 1.00 & 1.00 \\
		\midrule
		\multirow{3}{*}{Wine Quality}
		& 1x & 0.20 & 0.22 & 0.22 & 0.23 & 0.24 \\
		& 5x & 0.33 & 0.55 & 0.88 & 0.98 & 1.00 \\
		& 40x & 0.43 & 0.91 & 1.00 & 1.00 & 1.00 \\
		\bottomrule[1.5pt]
	\end{tabular}
		\caption{Percentage of extracted leaves with an
			informed attacker who knows a certain percentage of the 
			training
			data. Results are shown for different numbers of cover 
			queries that are sent between each attack query from the 
			tree extraction algorithm.}
		\label{tab:tabKnowledgeAttacker}
\end{table}

\vspace{-0.1em}
\section{Discussion}
\label{sec:discussion}

Adversarial machine learning and digital watermarking are vivid
research fields that have established a broad range of methods and
concepts. However, one can observe the following asymmetry: the former
community has focused on white-box attacks, while the latter has
extensively studied the black-box threat. Although recent research in
machine learning has started to also study black-box attacks more
thoroughly~\citep{PapMcDGoo+16,PapMcDGoo16b,TraZhaJuel+16}, existing
insights from digital watermarking potentially bring forth novel
ideas.  The other way round, we also show that knowledge from machine
learning can help to mitigate attacks against watermarks. %

The presented unified view opens interesting directions for future
work. The comparison of defenses between both fields in
Section~\ref{subsec:defenses}, for instance, discloses that stateful
detection strategies have not been considered in machine learning yet.
While we successfully transferred the security margin approach in this
paper, a line search detection based on a PCA, for example, could
further mitigate model-extraction attacks.  On the other side, the
various strategies, such as adaptive
re-learning~\citep{TraZhaJuel+16}, that have been used successfully
against classifiers mark a potential threat for watermark detectors as
well.

Furthermore, the watermarking community has conducted different
contests, where researchers could attack and defend watermarking
schemes under practical conditions, such as the ``Break Our
Watermarking System'' (BOWS) competition. These contests have promoted
a variety of publications that reveal shortcomings of existing
protection techniques and introduce novel
defenses~\mbox{\citep[e.g.][]
  {Wes06,ComPer07,CraJun07,Wes08,BasWes09,XieFurFon10}}, such as the
strong watermarking scheme \textit{Broken Arrows}~\citep{FurBas08}.
Based on our unified view, we encourage the organization of a similar
\emph{contest for adversarial machine learning}.  By imposing
researchers into the role of an attacker in a real scenario without
perfect knowledge, previously unknown questions and insights often come
to light.  \v{S}rndi\'{c} and Laskov~\citep{RndLas14}, for instance,
demonstrate the feasibility to evade a publicly available PDF malware
classifier with the insight that full knowledge of the classifier
features is not necessary. The contest could be structured similarly
to watermarking contests in different episodes, with each providing a
different level of knowledge about a defense or attack (see
Appendix~\ref{sec:appendixBOWS}).

Finally, we note that machine learning and watermarking are not the
only research areas that have to cope with an adversary. The identified
similarities between both research fields can be seen as part of a
bigger problem: \emph{Adversarial Signal
Processing}~\citep{Barni:2013aa}. More fields such as information or
multimedia forensics also deal with an adversary's presence and to our
knowledge the transfer of concepts between and to these areas has not
been addressed so far. %

\section{Conclusion}
\label{sec:conclusion}

Developing analysis methods for an adversarial environment is a
challenging task: First, these methods need to provide correct results
even if parts of their input are manipulated and, second, these
methods should protect from known as well as future attacks. The
research fields of adversarial learning and digital watermarking both
have tackled this challenge and developed a remarkable set of defenses
for operating in an adversarial environment.

In this paper, we show that both lines of research share similarities
which have been overlooked by previous work and enable transferring
concepts from one field to the other. By means of a systematization of
attacks, we are able to transform defenses for learning methods to the
domain of watermarking and vice versa. This not only opens new
perspective for designing joint defenses, but also allows for
combining techniques from both fields that have not been previously
coupled.

As part of our analysis, we identify interesting directions of future
research that enable the two communities to learn from each other and
combine the ``best of both worlds''. As one example of these
directions, we particularly encourage the organization of a public
competition for adversarial machine learning, where attacks and
defenses are put to the test in a competitive manner.

{\footnotesize \bibliographystyle{acm}
}

\appendix

\section{Linear Watermark Detector}\label{sec:appendixLinearSpecialCase}
\begin{figure}[htb]
	\centering
	\includegraphics[]{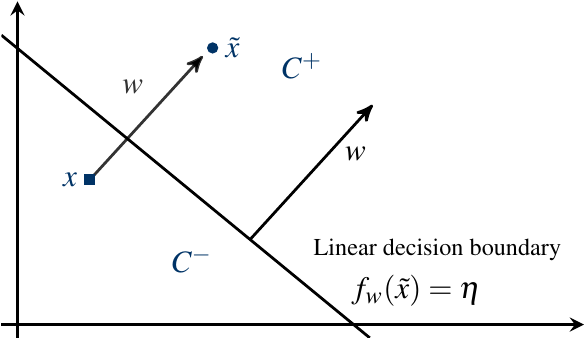}
	\caption{
		Geometrical view on the embedding and detection process of a
		simple watermarking scheme.
	}
	\label{fig:exampleFig}
\end{figure}

This section illustrates the watermarking process with a simple
watermarking scheme.
The process can be generally divided into two phases:
embedding and detection.
Let us, for instance, consider the \emph{additive spectrum
	watermarking scheme} which is also the embedding scheme used in the
image example from Figure~\ref{fig:realworldexamplewatermarking}.
In this scheme, the watermarked version $\wm$ of a signal $\x$ is
created by adding a watermarking vector $\w \in \mathbb{R}^{N}$ onto
$\x$ element-wise, that is,
\begin{align}
\wm = \x + \w \; .
\end{align}
The watermark $\w$ usually represents a random pattern.
In order to decide whether a signal contains the particular watermark,
a \emph{linear correlation detector} can be employed that uses the
following decision function
\begin{align}
f_{\w}(\wm) = \transp{\wm} \kern1pt \w \; .
\end{align}
The output is a weighted sum between $\wm$ and the watermark~$\w$. If
watermark and signal match, the correlation exceeds a pre-defined
threshold $\eta$. Geometrically, each signal corresponds to a point in
a vector space where the watermark describes a decision
boundary, as shown in Figure~\ref{fig:exampleFig}.
The result are two subspaces, one for the watermark's presence, one for
its absence. The detection thus works by determining which subspace an
input signal is currently in.

\section{Blind Newton Sensitivity Attack}\label{sec:bnsa_attack}
This section briefly recaps the Blind Newton Sensitivity Attack (BNSA)
that interprets the watermark removal as a non-linear optimization
problem~\citep{ComPerPer06}:
\begin{eqnarray}
\min & & d(t) \\
\text{subject to} & & f_{\w}(\wm') = f_{\w}(\wm+ t) = \eta.
\label{eq:BNSA2}
\end{eqnarray}
The objective function $d(t)$ %
measures the changes $t$ on the image $\wm$. For example, the
squared Euclidean norm $d(t) = \Vert t \Vert^2_2$ minimizes the length
of $t$ and therefore the necessary pixel changes.  At the same
time, the optimal solution must satisfy the constraint that the
detector does not detect the watermark.  A position on the boundary is
here sufficient, so that Equation~\eqref{eq:BNSA2} restricts the
decision function to $\eta$.

\begin{figure}
	\centering
	\includegraphics[]{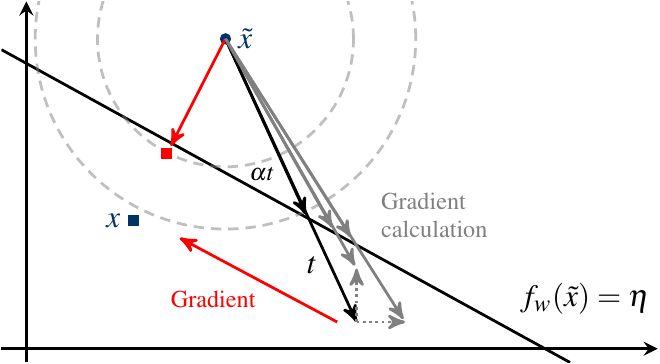}
	\caption{Blind Newton Sensitivity Attack. Queries around a boundary
		position reveal the function's gradient at this position to
		minimize the distance between the manipulated sample and the
		original one.}
	\label{fig:bnsa_attack}
\end{figure}

The adversary, however, does not know the function $f_{\w}(\wm')$,
since the watermark $\w$ is kept secret. Although only a binary 
detector output is observable, an attack is yet possible. To this end, 
Comesa\~na et al.\ rewrite the optimization problem into an 
unconstrained version:
\begin{eqnarray}
  \arg \underset{t \in \mathbb{R}^N}{\min} \; d(h(t)).
\label{eq:BNSA3}
\end{eqnarray}
The function $h(t)$ reflects the prior constraint by mapping $t$ to
the decision boundary. To this end, a bisection algorithm can be used
to find a scalar $\alpha$ such that $\alpha t$ lies on the decision
boundary. There is, however, another problem. As $h(t)$ has to map
each input vector to the boundary explicitly by running the bisection
algorithm respectively, a closed form to solve the problem is not
applicable. Therefore numeric iterative methods such as Newton's
method or gradient descent have to be used as
Figure~\ref{fig:bnsa_attack} exemplifies.

The attack starts with a random direction to locate the decision
boundary. After calculating an image at the boundary, it slightly
changes the vector at one position, maps the vector to the boundary
again and records the distance through this change. By repeating this
procedure for each feature direction, the attack is able to calculate
the gradient at this boundary position. This step yields the direction
in which the necessary changes $d(t)$ decreases fastest. In this way, 
the attack is able to locate a
boundary position that is closer to $\z$ than the previous position.
This process can be repeated, but in the case of a linear boundary,
the algorithm finishes after one iteration with the fewest necessary
changes.

In summary, the attack does not require a priori knowledge about the
detector's decision function and works only with a binary output.
Although the attack only converges to an optimal solution for convex
boundaries, it has been used against various watermarking schemes with
even polynomial and fractalized decision boundaries~\citep{ComPerPer06,
ComPer07, TonComPerBar15}.

\section{Security Margin Construction}\label{sec:appendixSecurityMargin}
The security margin's construction works as follows:
First, we choose a tree region and select the training data that
fall inside this particular region.
Next, we estimate the distribution of the selected training data at
each dimension through a kernel-density estimation.
In this way, no a priori assumptions about their distribution are
required. Finally, the distribution in each dimension is used to define
the margin at the boundary in this dimension. To this end, we set
the margin to the feature value where the probability of occurrence is
smaller than a certain threshold. In Figure~\ref{fig:DWtoMLtree}(a),
for example, the top right tree regions has a smaller security margin,
since more training data are near the boundary. On the contrary, the
most left region exhibits fewer training samples near the boundary, so
that a larger margin can be defined. By defining the security margin in
this statistical way, we can control the false alarm rate that a
honest query falls inside the margin. We repeat the process for each
tree region.

\section{BOWS Contest}\label{sec:appendixBOWS}

``Break Our Watermarking System'' or BOWS is a contest that has been
held twice in the watermarking community. The
latest contest is divided into three subsequent episodes, where only
the last episode reveals the underlying watermarking scheme.

\vspace{3pt}
\begin{enumerate}
\setlength{\itemsep}{4pt}
\item At the beginning of the contest, 3 watermarked images are
  available together with an online watermark detector that allows
  30~calls per day.  This episode models an attacker with limited
  knowledge and capabilities. The participants are required to operate
  with few queries and need to carefully construct their attacks.

\item In the next episode, the daily rate limited is dropped and the
  participants can perform different forms of oracle and
  watermark-estimation attacks against the detector. The episode
  models a stronger attacker, yet only 3 images are available for
  inferring the pattern embedded by the watermarking scheme.

\item Finally, the same watermark is embedded into 10,000~images and
  the underlying watermarking scheme is released. This episode models
  a very strong adversary with full knowledge of the scheme together
  with access to a large set of images. Ultimately, a watermarking
  scheme should remain secure even in this setting.
\end{enumerate}
\vspace{3pt}

For each episode, a hall of fame on the respective website documents
the participant's success regarding the image quality and the launched
attacks. Further information on the design of both contests are provided
by Piva and Barni~\citep{PivBar07} as well as Furon and
Bas~\citep{FurBas08}, respectively.
\vspace{3pt}

\end{document}